\documentclass[%
 amsmath,amssymb,
 twocolumn,
 aps,
  prl,
  superscriptaddress,
float,
floatfix,
]{revtex4-1}

\usepackage{graphicx}
\usepackage{dcolumn}
\usepackage{bm}

\usepackage{times}

\begin{document}

\preprint{APS/123-QED}

\title{Demonstration of a time scale based on a stable optical carrier}

\author{William R. Milner}
\email{william.milner@colorado.edu}
\affiliation{JILA, NIST and University of Colorado, 440 UCB, Boulder, Colorado 80309, USA}
\author{John M. Robinson}
\affiliation{JILA, NIST and University of Colorado, 440 UCB, Boulder, Colorado 80309, USA}
\author{Colin J. Kennedy}
\affiliation{JILA, NIST and University of Colorado, 440 UCB, Boulder, Colorado 80309, USA}
\author{Tobias Bothwell}
\affiliation{JILA, NIST and University of Colorado, 440 UCB, Boulder, Colorado 80309, USA}
\author{\\Dhruv Kedar}
\affiliation{JILA, NIST and University of Colorado, 440 UCB, Boulder, Colorado 80309, USA}
\author{Dan G. Matei}
\affiliation{Physikalisch-Technische Bundesanstalt, Bundesallee 100, 38116 Braunschweig, Germany}
\author{Thomas Legero}
\affiliation{Physikalisch-Technische Bundesanstalt, Bundesallee 100, 38116 Braunschweig, Germany}
\author{Uwe Sterr}
\affiliation{Physikalisch-Technische Bundesanstalt, Bundesallee 100, 38116 Braunschweig, Germany}
\author{Fritz Riehle}
\affiliation{Physikalisch-Technische Bundesanstalt, Bundesallee 100, 38116 Braunschweig, Germany}
\author{\\ Holly Leopardi}
\affiliation{National Institute of Standards and Technology, 325 Broadway, Boulder, Colorado 80305, USA }
\author{Tara M. Fortier}
\affiliation{National Institute of Standards and Technology, 325 Broadway, Boulder, Colorado 80305, USA }
\author{Jeffrey A. Sherman}
\affiliation{National Institute of Standards and Technology, 325 Broadway, Boulder, Colorado 80305, USA }
\author{Judah Levine}
\affiliation{National Institute of Standards and Technology, 325 Broadway, Boulder, Colorado 80305, USA }
\author{Jian Yao}
\email{jian.yao@colorado.edu}
\affiliation{National Institute of Standards and Technology, 325 Broadway, Boulder, Colorado 80305, USA }
\author{Jun Ye}
\email{Ye@jila.colorado.edu}
\affiliation{JILA, NIST and University of Colorado, 440 UCB, Boulder, Colorado 80309, USA}
\author{Eric Oelker}
\email{ericoelker@gmail.com}
\affiliation{JILA, NIST and University of Colorado, 440 UCB, Boulder, Colorado 80309, USA}


\begin{abstract}
We demonstrate a time scale based on a phase stable optical carrier that accumulates an estimated time error of $48\pm94$ ps over 34 days of operation. This all-optical time scale is formed with a cryogenic silicon cavity exhibiting improved long-term stability and an accurate $^{87}$Sr lattice clock.  We show that this new time scale architecture outperforms existing microwave time scales, even when they are steered to optical frequency standards. Our analysis indicates that this time scale is capable of reaching a stability below $1\times10^{-17}$ after a few months of averaging, making timekeeping at the $10^{-18}$ level a realistic prospect.

\end{abstract}

\pacs{Valid PACS appear here}
\maketitle



 
 Accurate and precise timing is critical for a wide array of applications, ranging from navigation and geodesy to studies of fundamental physics~\cite{Ashby2003, domain, Wolf_PRL, takano2016geopotential, lombardi2016accurate, Kennedy}.  The worldwide time standard, Coordinated Universal Time (UTC), is synthesized from a global network of atomic clocks and disseminated at monthly intervals.  National metrology institutes bridge the gap between updates of UTC by broadcasting independent time scales derived from ensembles of microwave local oscillators steered to accurate atomic frequency standards~\cite{UTC(PTB), UTC(OP)}.  To advance the frontier of precision timekeeping, the development of both improved local oscillators and atomic frequency standards is imperative. 
 
 
 

Optical atomic clocks, orders of magnitude more accurate and stable than their microwave counterparts~\cite{Bloom, Travis, Ludlow, takano2016geopotential, PTB_Accuracy, Al_Accuracy}, show promise as frequency standards for time scale applications.  Recent efforts to incorporate optical clocks into existing microwave timescales have lead to improved performance~\cite{NICT, PTBTimescale, yao2019optical}. However, despite the fact that optical clocks have demonstrated mid-$10^{-17}$ level stability in one second of averaging~\cite{Eric,YbZDT}, time scales steered to optical standards have thus far required weeks of averaging to reach $10^{-16}$ level precision~\cite{yao2018, yao2019optical}. This disparity in performance arises due to down conversion of noise from the local oscillator -- a consequence of steering to an atomic standard in the presence of dead time -- which degrades the long-term stability of the time scale~\cite{yao2019optical}.  This limitation motivates the development of local oscillators with improved stability, particularly at averaging times around the typical interval between clock measurements ($10^3$ to $10^5$ s). In parallel, improvements in local oscillator stability allow a timescale to maintain a competitive level of performance even when relaxing the requirements on optical clock uptime.

In this Letter, we report on the first realization of an all-optical time scale that outperforms state-of-the-art microwave oscillators steered to either microwave or optical frequency standards. This time scale consists of an optical local oscillator (OLO) based on a cryogenic silicon reference cavity which is steered daily to an accurate $^{87}$Sr lattice clock \cite{Toby} over a month-long campaign. During this period, the frequency stability of the OLO surpasses that of the hydrogen masers in the UTC(NIST) time scale at all averaging intervals up to multiple days~\cite{Supplemental}, demonstrating the requisite stability for improved time scale performance. Our analysis indicates that daily steering of the OLO frequency with 50\% clock uptime allows for a time scale instability below the $10^{-17}$ level within 85 days of operation. Our local oscillator frequency is easily predictable using conventional time scale steering algorithms, allowing us to limit the estimated time error to only $48\pm94$ ps after $34$ days of operation.  The continuous availability of the OLO coupled with the on-demand performance of our optical clock make our system viable for future inclusion in UTC(NIST).  This new variant of time scale harnesses both the improved accuracy and stability of optical standards and provides a viable blueprint for the upgrade of time scales worldwide.


After a decade of development~\cite{kessler2012sub,zhang2017ultrastable}, cryogenic silicon reference cavities are now a proven platform for laser stabilization at the mid-$10^{-17}$ level~\cite{John, PTB_cavity}.  The exceptional short-term stability of these local oscillators has enabled advances in optical clock stability~\cite{Eric}. These systems outperform all free-running local oscillators at averaging times below $1 \times 10^{4}$ seconds~\cite{Eric} and exhibit orders-of-magnitude lower frequency drift than other OLOs~\cite{PTB_drift,John}.  However, achieving a stability commensurate with the best microwave oscillators at longer averaging times has remained an elusive goal, hampering their usefulness as time scale flywheel oscillators. The OLO used in our time scale, based on a $21$ cm long Si cavity operating at $124$ K, was recently optimized to significantly improve its long-term stability. 
The use of super-polished optics and thermal control of the environment limit parasitic etalons and active optical power stabilization reduces frequency excursions from laser intensity drift~\cite{Supplemental}. 

We combine our local oscillator with an accurate optical frequency standard to form an all-optical time scale. Over a 34 day interval, a strontium lattice clock with systematic uncertainty of $2.0 \times 10^{-18}$ ~\cite{Toby} is used to track the OLO frequency with 25 percent uptime.  Daily measurements of the OLO allow us to build a reliable predictive model of its frequency evolution.  As new frequency data become available, the model is updated to better reflect its current behavior.  The OLO is steered using the model to correct for changes in its frequency over time, and any residual frequency fluctuations ultimately determine the time scale stability.  The analysis required to realize the time scale was carried out in post-processing, though we emphasize that our approach is compatible with real-time implementation.

To track frequency excursions larger than the low-$10^{-16}$ level during intervals when the optical clock is offline, the OLO is compared with two independent ultrastable lasers based on a $6$ cm silicon cavity operated at $4$ K~\cite{John} and a $40$ cm ultra-low expansion (ULE) cavity~\cite{Bishof}. Because the three systems have comparable short-term stability, one may use a three-cornered hat analysis to identify any significant frequency jumps in the OLO and update the predictive model accordingly~\cite{Supplemental}.  

 
\begin{figure}
\includegraphics[width=0.47\textwidth]{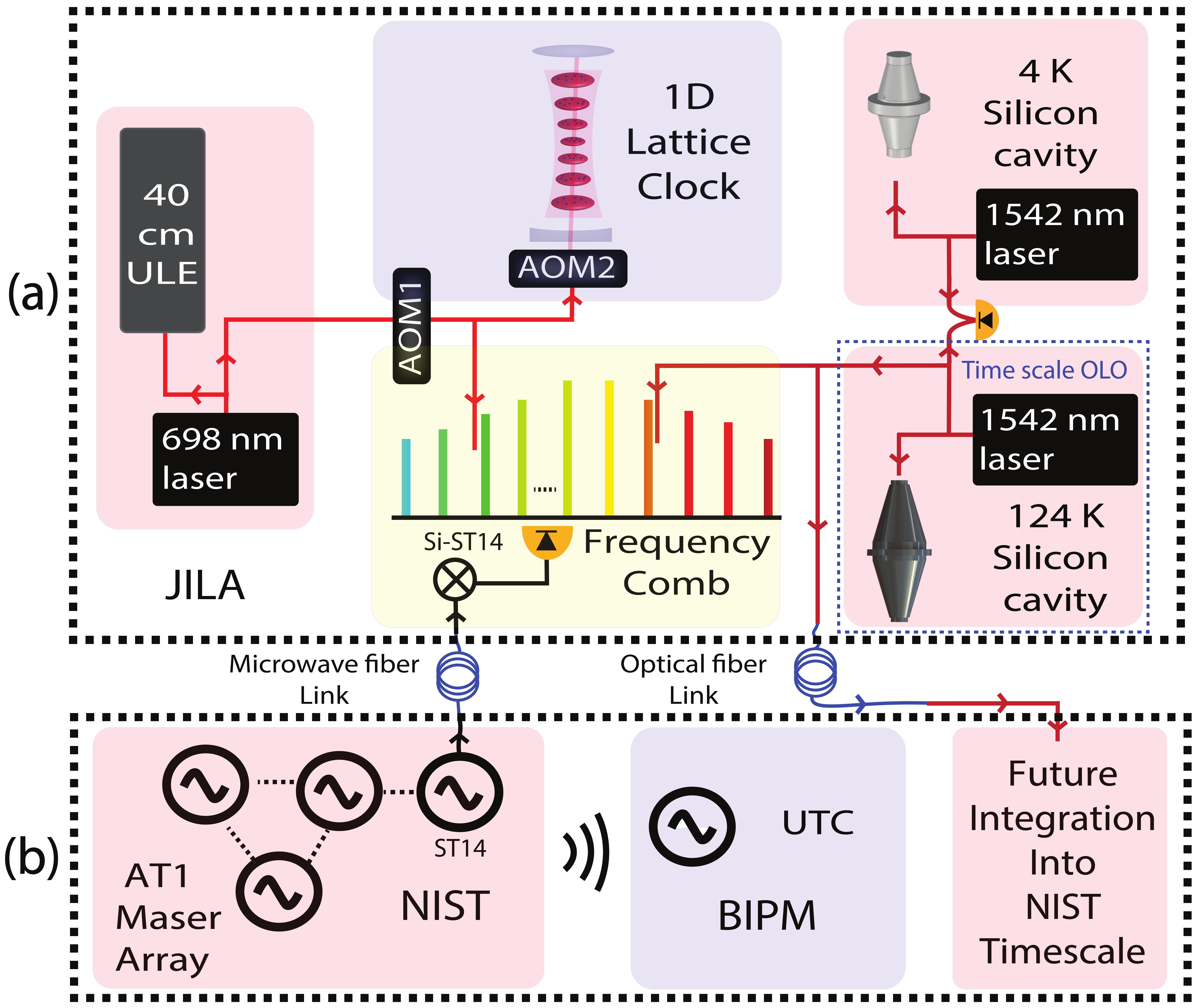}
\caption{Schematic of the optical time scale. (a) An array of three lasers are locked to ultrastable Fabry-P\'{e}rot resonators. A femtosecond frequency comb transfers the stability of the OLO ($124$ K Si cavity) from $1542$ nm to a prestabilized laser at $698$ nm used to perform clock spectroscopy in a 1D $^{87}$Sr lattice clock.  (b) AT1, a free running microwave time scale at NIST is compared continuously against the OLO signal over a fiber optic link using a hydrogen maser (ST14) as a transfer oscillator.  An optical fiber link between JILA and NIST allows for stable transfer of the optical time scale to NIST for future integration into UTC(NIST).}
\label{Figure1}
\end{figure}

\begin{figure}
 
\includegraphics[width=3.375in]{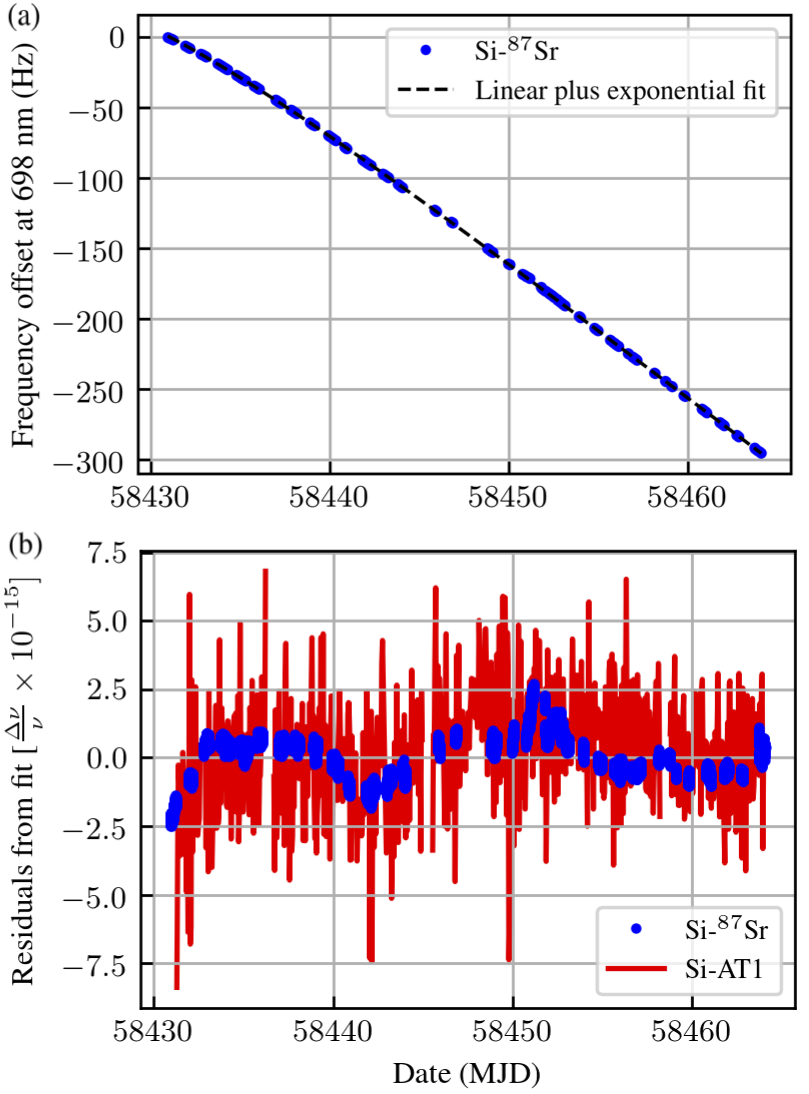}
\caption{Frequency record of the local oscillator. (a) The OLO frequency (Si) is measured at 698 nm using a $^{87}$Sr lattice clock. A linear plus exponential trend, $a + b t + c e^{-\frac{t}{d}}$, agrees well with the raw frequency data. The fit parameters are $a =24.16$ Hz, $b = -9.632$ Hz$/$day, $c = -23.17$ Hz, and $d = 7.813$ days.  (b) The residuals of the OLO comparisons against the $^{87}$Sr clock and the NIST AT1 time scale after subtracting the drift trend from (a) from both datasets.}
\label{Figure2}
\end{figure}

A schematic of our optical time scale is presented in Fig. 1(a). In order to reference the $^{87}$Sr clock laser to the 124 K silicon cavity, we transfer its optical stability from 1542 nm to a prestabilized laser at 698 nm using a femtosecond Er:fiber frequency comb with negligible additive instability~\cite{Eric}.  The frequency corrections applied to AOM1 by the stability transfer servo are recorded to monitor the relative frequency fluctuations between the 40 cm ULE cavity and the OLO. The stabilized 698 nm light is then tuned to resonance for the $^{87}$Sr clock transition using AOM2. The AOM2 correction signal is recorded and yields the OLO frequency relative to the $^{87}$Sr transition.  An optical beatnote at 1542 nm between the OLO and the 6 cm Si cavity serves as a continuous monitor of their frequency difference.  Fig. 1(b) depicts AT1, a free running microwave time scale at NIST. Using a hydrogen maser as a transfer oscillator, AT1 is compared remotely with the local oscillator over a stabilized fiber-optic link~\cite{Supplemental}. To enable this comparison, the OLO is down converted to the RF domain using a frequency comb.  This provides an additional record of the long-term performance of the OLO that is nearly continuous (95\% uptime) over the measurement campaign.  We note that AT1 is chosen rather than UTC(NIST) due to its superior stability over the averaging intervals of relevance to this study.




A record of the OLO frequency during the data campaign spanning from a modified Julian date (MJD) of $58430$ to $58464$ is presented in Fig. 2(a).  The clock ran daily with the exception of MJD 58444 and 58447.  Three days before the first measurement, the optical power incident on the cavity was changed to reset an intensity noise servo. Consistent with prior silicon cavity drift studies, the frequency evolution of the OLO after adjusting the incident optical power is well modeled by a constant linear drift plus an exponential relaxation term: $a + b t + c e^{-\frac{t}{d}}$ \cite{John}.   Fig. 2(b) shows the residuals of the OLO comparisons with the clock and AT1 after subtracting the modeled drift trend determined by a fit to the $^{87}$Sr clock data.  Perfect correlation between the two data sets is not expected as both AT1 and the microwave link contribute additional instability to the Si-AT1 record~\cite{Supplemental}. 


During the interval between clock operation on MJD 58441 and 58442, two frequency jumps on the OLO were identified with a combined amplitude of $5.02 \times 10^{-15}$. A correction of the same magnitude is applied to all data after this step when performing the analysis presented in this work. No significant change in the long-term drift trend of the local oscillator was observed following these excursions~\cite{Supplemental}. 


\begin{figure}
\includegraphics[width=3.25in]{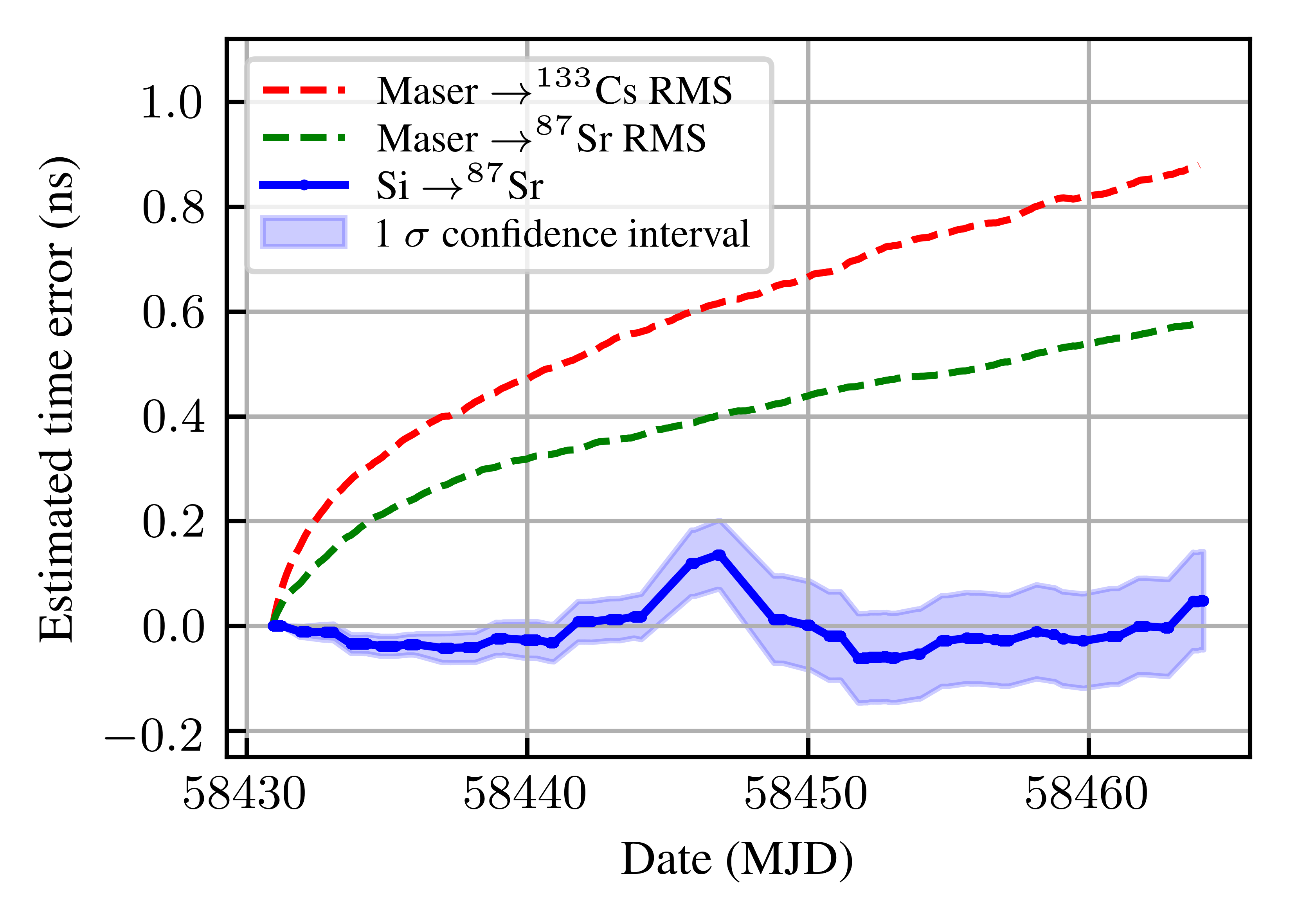}
\caption{An estimate of the time error evolution of the optical time scale over the $34$ day data campaign results in an integrated value of $48\pm94$ ps. The peak-to-peak value of $197$ ps is dominated by a four day window that includes the two days when the $^{87}$Sr clock was not operated. The RMS spread in time error for two time scales based on repeated simulations of a maser steered to either a microwave or optical frequency standard are shown for comparison~\cite{Supplemental}.} 
\label{Figure3}
\end{figure}

To realize a time scale, the OLO frequency record in Fig 2(a) is steered using a predictive model to minimize its offset from the atomic frequency standard.  The predictive model utilizes a Kalman filter to estimate the frequency of the OLO at a given time based on prior measurements with the clock.  Kalman filtering techniques are commonly used in time scales to model the frequency of hydrogen masers~\cite{Levine, JY1}. These models approximate the hydrogen maser as a linearly drifting oscillator and update the model parameters as new frequency measurements arrive.  The drift in the OLO frequency between daily measurements can be modeled using a quadratic function: $k_{0} + k_{1}t + \frac{k_{2}t^{2}}{2}$ and traditional Kalman filtering techniques are applicable. The model prediction is determined by a state vector $[k_{0}, k_{1}, k_{2}]$ that is updated epoch-by-epoch when the $^{87}$Sr clock is running.  Further detail on the Kalman filter algorithm is provided in~\cite{Supplemental}.

To evaluate the performance of a time scale, one typically compares it against a reference time scale with significantly lower timing uncertainty.  To our knowledge, no such time scale exists in this case.  Instead we treat the $^{87}$Sr clock as an ideal frequency reference and examine the fractional frequency offset between the steered OLO record and the clock transition frequency, hereafter referred to as the prediction error. We define the time error of our time scale as the integral of the prediction error over time.

If the frequency record were continuous, the time error could be determined to within the measurement precision of the clock.  However, a finite gap of time separates the frequency measurements in Fig. 2(a), ranging from the $5$ second interrogation cycle of our experiment to $24$ hours between daily measurements.  Most of the time error accumulates during the longer gaps, when the Kalman filter must accurately predict changes in the OLO frequency without new measurement data from the clock.  The time error contribution from a gap is simply the gap duration multiplied by the mean prediction error during this interval.  However, the latter quantity cannot be determined exactly from the available data.  Instead, we estimate the mean prediction error by averaging the values before and after the gap and multiply by the gap duration to compute an estimated time error.  We compute a $1\sigma$ confidence interval for the estimated time error through repeated simulations of the OLO frequency during each gap to determine the uncertainty in the estimation of the prediction error~\cite{Supplemental}.


An estimate of the integrated time error of our optical time scale is presented in Fig. 3. After 34 days of integration our all-optical time scale accumulates an error of $48\pm 94$ ps. For comparison we simulate time scales consisting of a hydrogen maser steered to a $^{133}$Cs fountain for 24 hours/day and a hydrogen maser steered to a $^{87}$Sr optical clock for 6 hours/day using the same Kalman filter and noise models for the maser and fountain described in~\cite{yao2018}.  The typical performance of both time scales is assessed by computing time errors from repeated simulations\cite{Supplemental}, and their RMS spread over a 34 window is depicted in Fig. 3. Both exhibit a larger time error than the all-optical time scale. 

\begin{figure}
\includegraphics[width=3.37in]{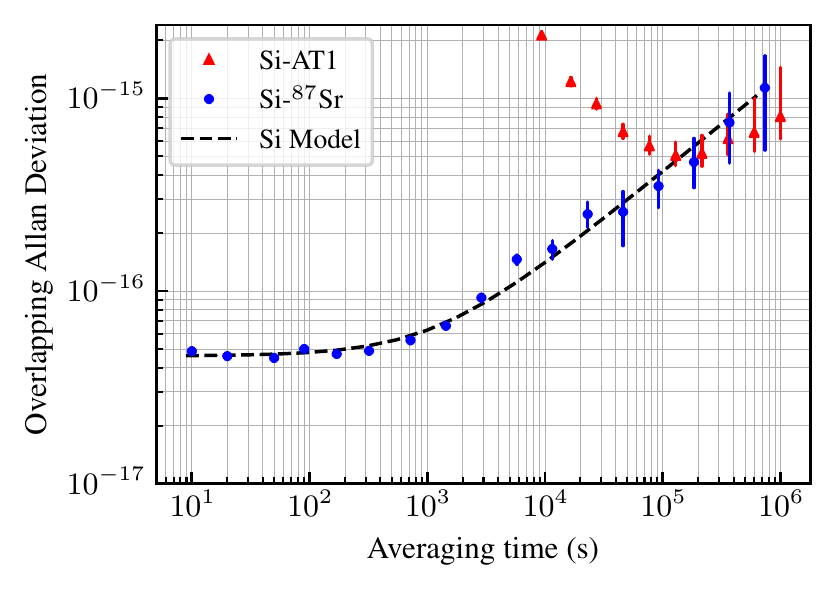}
\caption{The stability of our OLO is determined from daily measurements with the $^{87}$Sr clock. The silicon cavity stability is computed from the detrended $^{87}$Sr data in Fig 2(b) using a gap-tolerant Allan variance similar to \cite{sesia2008estimating}.  The data is fit to a noise model with an instability of $\sigma = 1.3\times$ $10^{-18}\sqrt{\tau\textrm{(s)}}$ at long averaging times. The long-term stability of the OLO is also inferred from a continuous measurement against the NIST AT1 time scale.}
\label{Figure4}
\end{figure}


Because the optical clock is run intermittently, the long-term stability of the time scale will be limited by a $1/\sqrt{\tau(s)}$ slope arising from aliased local oscillator noise akin to the Dick effect~\cite{yao2018,yao2019optical}.  Determining this stability limit requires an accurate characterization of the OLO.  We evaluate the stability of our OLO by analyzing the frequency noise of the residuals in Fig. 2(b). One complicating factor are the gaps in the frequency record during clock downtime. A gap-tolerant Allan variance similar to \cite{sesia2008estimating} is used to compute an estimated stability of the OLO out to multiple day averaging intervals. 

The result of this analysis is plotted in Fig. 4.  The OLO stability is fit to a noise model that includes the known thermal noise floor~\cite{Eric} and a random walk frequency noise term, resulting in an instability at long averaging times consistent with $\sigma_{RW} = 1.3\times$ $10^{-18}\sqrt{\tau\textrm{(s)}}$.  The OLO maintains an instability below $10^{-15}$ out to $6\times10^{5}$ s, more than an order of magnitude improvement over the previous characterization of this system~\cite{PTB_cavity}. The frequency stability of the Si-AT1 record is presented as well and its value at averaging times past $10^5$ s agrees with the clock measurement within statistical uncertainty.  At shorter averaging times, the stability is consistent with a noise model~\cite{Supplemental} accounting for instability from the microwave link, the OLO, and AT1~\cite{McGrew}.

With an accurate noise model for the OLO in hand, we now consider the anticipated long-term stability of our time scale as a function of optical clock duty-cycle. Similar to~\cite{yao2019optical,yao2018}, we simulate a lengthy local oscillator frequency record using the model presented in Fig. 4 with the drift trend from Fig. 2(a) added.  This record is then steered to a simulated $^{87}$Sr lattice clock for a fixed interval each day using the same Kalman filtering techniques described above.  We compute an Allan deviation of the prediction error to determine the stability of the time scale.  To quantify the impact of our improved local oscillator we carry out the same analysis for a similar time scale where the OLO has been substituted with a hydrogen maser.  The noise model for the simulated hydrogen maser is based on the typical stability of the best performing masers in the UTC(NIST) time scale~\cite{yao2018}.

Fig. 5 shows the results of our analysis.  As anticipated, the long-term stability of the time scale improves with increased clock uptime and reduced local oscillator noise and is reasonably consistent with the expected instability limit from aliased local oscillator noise past $10^6$ s~\cite{Supplemental}.  When the optical clock is run with the same duty cycle, the steered OLO significantly outperforms a steered hydrogen maser at all averaging times.   Even when steering one hour per day, our time scale is more stable than a hydrogen maser steered with a 50 percent duty cycle. This capability allows for competitive time scale performance with significantly relaxed uptime requirements.  Based on this analysis, we expect a stability of approximately $1.8 \times 10^{-17}$ after a 34 day campaign with an average clock uptime of 6 hours/day. This is in good agreement with the observed integrated time error of $48\pm94$ ps over 34 days, or $1.6\pm3.2 \times 10^{-17}$ in fractional units.  When operating the clock 12 hours per day, our all-optical time scale remains at or below the $2 \times 10^{-16}$ level at all averaging times and is projected to reach a stability below $10^{-17}$ after only 85 days of operation.  Additional effort on automation should allow for a clock duty cycle well above $50\%$.  Using an array of $N$ independent silicon cavities would improve the stability by a factor of 1/$\sqrt{N}$~\cite{yao2019optical}. 

\begin{figure}
\includegraphics[width=3.375in]{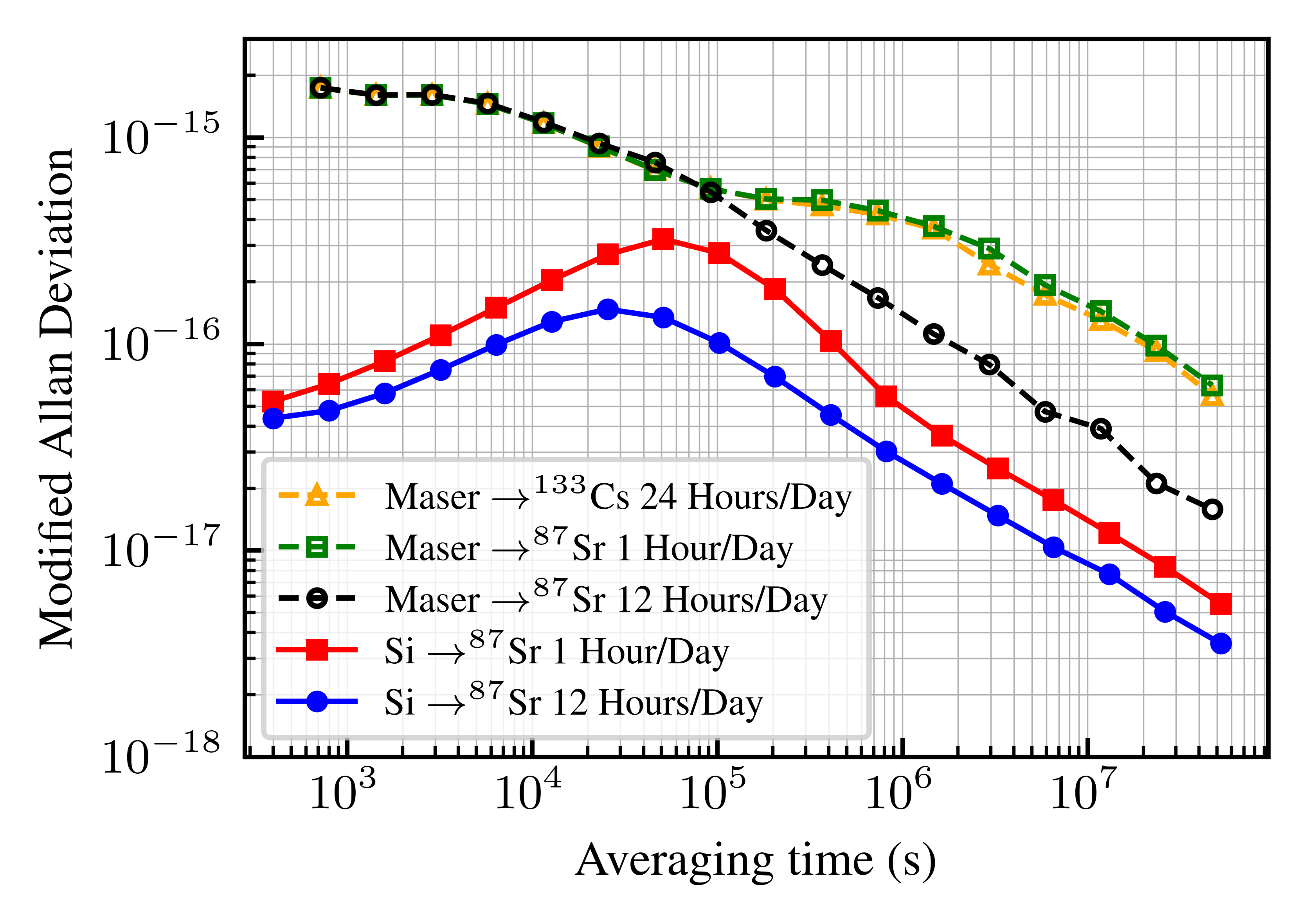}
\caption{Expected fractional frequency stability of the optical time scale. The stability of our optical time scale is analyzed for two optical clock duty cycles.  Our optical time scale is compared to a hydrogen maser based time scale steered to an optical lattice clock with identical uptime or a cesium fountain clock operating continuously.} 
\label{Figure5}
\end{figure}

By combining an improved local oscillator with an accurate high-uptime optical clock, we have demonstrated a novel time scale architecture with enhanced stability. Additional technical upgrades of our silicon cavity can further improve our optical time scale stability, including greater passive thermal isolation, shorter optical path lengths and operation closer to the silicon coefficient of thermal expansion zero crossing.  In addition, reducing the optical power incident on the cavity offers the capability to reduce the linear drift~\cite{John}.  

Future efforts will leverage existing time transfer infrastructure in Boulder, CO to incorporate this optical technology into the UTC(NIST) time scale. An underground fiber network is in place to support phase-stabilized optical signal transfer from JILA to NIST with negligible excess noise~\cite{foreman2007, Seth_Foreman_PRL}. Using a femtosecond frequency comb~\cite{Leopardi,Fortier06}, our optical time scale signal will be linked to UTC.

\vspace{4mm} 

\begin{acknowledgments}
We thank J.A. Muniz, T.R. O'Brian, A. Bauch, and J.L. Hall for careful reading of the manuscript and L. Sonderhouse for technical contributions. This work is supported by the National Institute of Standards and Technology, Defense Advanced Research Projects Agency, Air Force Office for Scientific Research, National Science Foundation (NSF PHY-1734006), Physikalisch-Technische Bundesanstalt, and the Cluster of Excellence (EXC 2132 Quantum Frontiers).  U.S. and T.L. acknowledge support from the Quantum sensors (Q-SENSE) project supported by the European Commission’s H2020 Marie Skłodowska-Curie Actions Research and Innovation Staff Exchange (MSCA RISE) under Grant Agreement Number 69115.  E.O. and C.J.K acknowledge support from the National Research Council postdoctoral fellowship. 

\vspace{0.5mm}
\noindent W.R.M. and E.O. contributed equally to this work.
\vspace{1.0mm}

\end{acknowledgments}

\bibliography{main}

\end{document}


\preprint{APS/123-QED}

\title{Supplemental material to \\Demonstration of a time scale based on a stable optical carrier}

\pacs{Valid PACS appear here}
\maketitle
\section{Continuous monitoring of the local oscillator frequency}

\begin{figure}[b]

\includegraphics[width=3.37in]{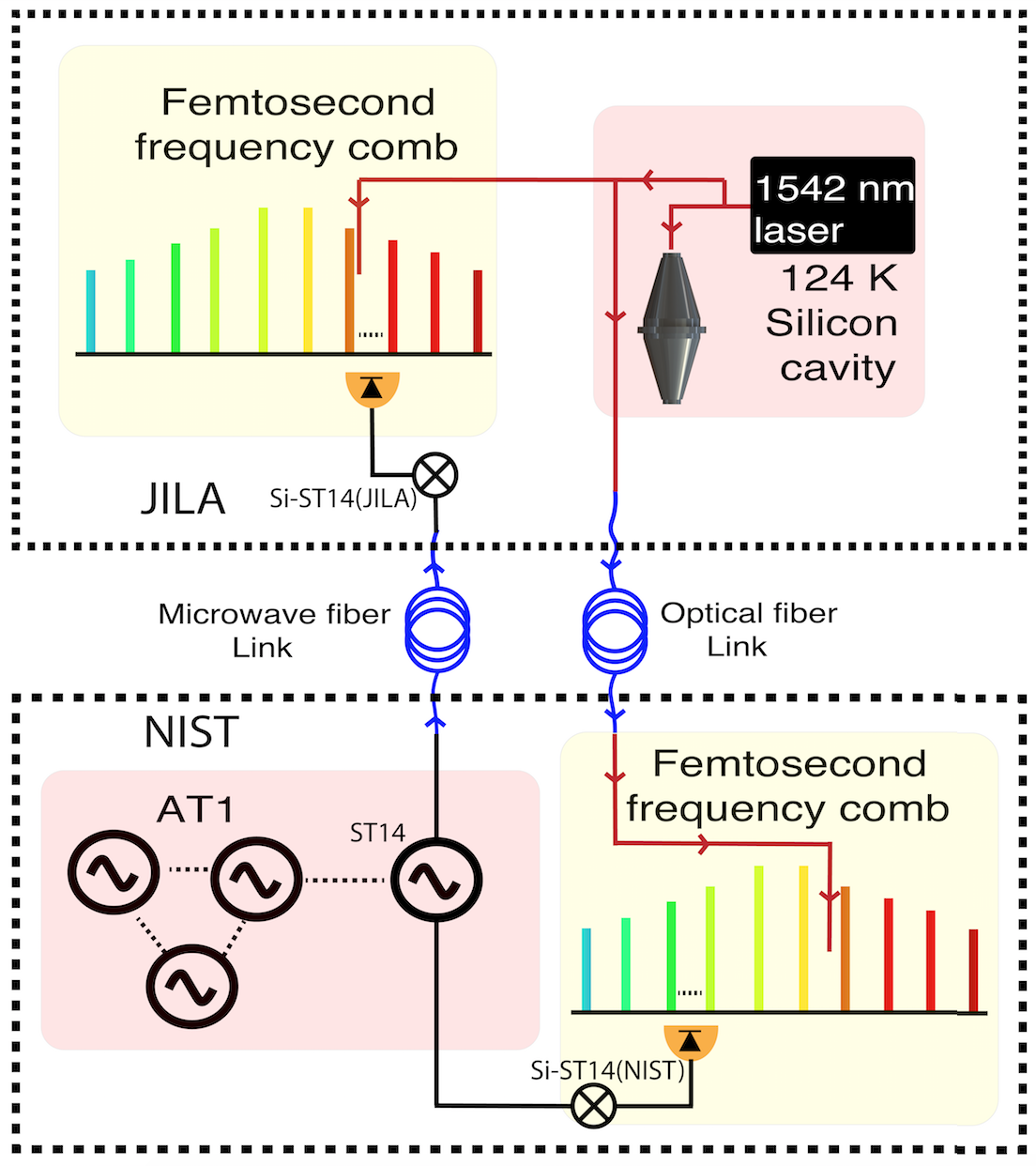}
\caption{Schematic of the frequency comparison measurement between the OLO and the NIST AT1 time scale.  A hydrogen maser, ST14, is used as a transfer oscillator.  ST14 is compared locally with AT1 and remotely with the OLO using a stabilized fiber optic link and the Si-AT1 signal is computed from the two measurements. The OLO is down-converted to the microwave domain using a frequency comb. The analysis  presented in Fig. 2(b) and Fig. 4 is generated by data from the frequency comb located at JILA. To characterize the noise added by the microwave link, a local measurement of Si-ST14 is performed at NIST over 5 days using a second frequency comb.  This utilizes an existing phase-stabilized optical fiber link which adds negligible instability.}
\label{SupplementSchematic}
\end{figure}

To supplement the intermittent frequency measurements of our optical local oscillator (OLO) with the $^{87}$Sr lattice clock, the OLO frequency was also compared continuously with AT1, a free running microwave time scale operating at NIST.  AT1 is chosen rather than UTC(NIST) because of its improved stability at averaging times below $10^6$ s.  To facilitate this comparison, the OLO frequency is down converted to the RF domain using a self-referenced Er:fiber frequency comb.  The frequency difference ($f_\textrm{beat}$) between the OLO and the nearest comb tooth is stabilized to 35 MHz by actuating on the comb repetition rate ($f_\textrm{rep} \approx 250$ MHz) as described by the relation: 
$f_\textrm{beat} = f_\textrm{Si} - n f_\textrm{rep} - f_\textrm{ceo}$.  Here, $f_\textrm{Si}$ is the OLO frequency, $n$ is an integer, and $f_\textrm{ceo}$ is stabilized to 35 MHz using an f-2f interferometer.  The repetition rate, $f_\textrm{rep}$ inherits the stability of the OLO and is used to synthesize an RF signal by measuring the optical beat between adjacent comb teeth.  This signal is mixed down to 1 MHz using a RF signal synthesized from a NIST hydrogen maser, hereafter referred to as ST14, and counted with a 1 s gate time using a zero-dead-time lambda-type frequency counter.  This allows us to continuously monitor $y_\textrm{Si-ST14}$, the fractional frequency difference between the OLO and ST14.  The fractional frequency difference between ST14 and the AT1 timescale ($y_{\textrm{ST14-AT1}}$) is measured in tandem at NIST and we use both to compute the Si-AT1 beat as follows: $y_\textrm{Si-AT1} = y_\textrm{Si-ST14} + y_\textrm{ST14-AT1}$.  Because the NIST time scale measurement system (TSMS) records $y_\textrm{ST14-AT1}$ in 720 s intervals, the Si-ST14 record is decimated in three steps by a factor of 720 and interpolated to the TSMS time stamps prior to computing $y_\textrm{Si-AT1}$.  To transfer ST14 from NIST to JILA, a 1 GHz signal synthesized from ST14 is used to amplitude modulate a $1.3$ micron wavelength laser which is transmitted over a group delay stabilized fiber-optic link~\cite{campbell2008absolute}.  Over the 34 day data campaign, the microwave fiber link and frequency comb located at JILA maintained a 95\% uptime to determine the $y_\textrm{Si-AT1}$ beat. 

We note that drawing conclusions about the performance of the OLO from the Si-AT1 record is more nuanced than evaluating its stability directly using an optical clock.  In this case, both the reference (AT1) and the link used to transfer microwave signals from NIST to JILA contribute appreciable instability to $y_\textrm{Si-AT1}$.  Interpreting the stability of $y_\textrm{Si-AT1}$ requires comprehensive knowledge of these contributions.  To estimate the instability added by AT1, we use the measurement of AT1 presented in~\cite{McGrew}.  The instability added by the microwave link is determined by performing a differential measurement.  Over a 5 day interval during the data campaign (MJD 58438-58442), the OLO was transferred to NIST over a phase stabilized optical fiber link and measured against ST14 locally using a second Er:fiber frequency comb~\cite{leopardi2017single}. Because the optical link adds negligible noise to the OLO~\cite{foreman2007}, we may estimate the noise added by the microwave link by examining the difference between these frequency records:  $y_\textrm{link} = y_\textrm{Si-ST14(JILA)} - y_\textrm{Si-ST14(NIST)}$.

\begin{figure}[t]

\includegraphics[width=3.37in]{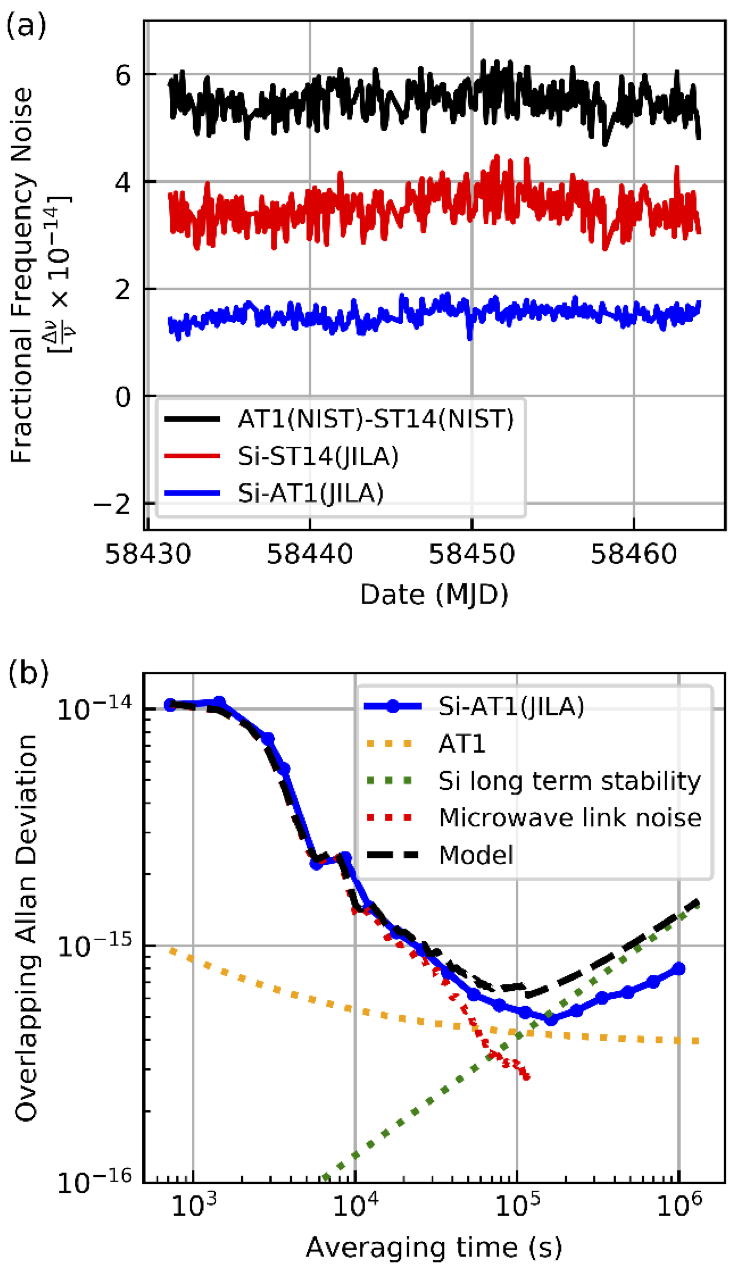}
\caption{Frequency comparison between the OLO and the NIST AT1 time scale. (a) Time series of the beats between the various oscillators highlighting the improved stability of the OLO.  The traces are labeled to indicate that they were measured either locally at NIST or remotely at JILA and are offset for clarity. (b) Fractional frequency stability of the Si-AT1(JILA) record. The data is compared against a model including contributions from AT1, microwave link noise, and the OLO model from Fig. 4.}
\label{AT1Comparison}
\end{figure}

Fig. ~\ref{AT1Comparison} depicts the time series of the three microwave signals after subtracting the linear frequency drift of ST14 and the OLO drift trend from Fig. 2(a).  Each signal has been decimated by an additional factor of 10 so that each is dominated by the long term performance of the oscillators rather than short term microwave link noise.  This plot highlights the improved stability of the OLO in comparison with ST14.

The long term stability of the OLO can be inferred from the Allan deviation of the Si-AT1(JILA) record in Fig. ~\ref{AT1Comparison}. The plot also includes a model of the anticipated stability where the measured link noise, a noise model of AT1, and the OLO model from Fig. 4 are added in quadrature.  Because of the shorter duration of the microwave link noise measurement, its magnitude is only known out to $10^5$ s and its contribution to the model is set to zero at longer averaging times.  At averaging times less than $10^{5}$ seconds, the data is in good agreement with the model and is slightly below the model prediction at longer averaging times. This data suggests that the intermittent measurements of the OLO with the Sr clock provide an accurate picture of its true long-term performance.

\section{Kalman filter steering algorithm for an all-optical time scale}


To minimize the time error during periods when an accurate frequency reference is unavailable, a time scale local oscillator must have a predictable frequency. The frequency of the local oscillator can then be steered using a predictive model to keep it as close to the value of the frequency reference as possible.  Kalman filtering techniques are often used to construct such a model based on periodic measurements of the local oscillator frequency against an atomic frequency reference~\cite{yao2018,yao2019optical}.  For a hydrogen maser, a linear model of the form $f(t+\Delta t) = \hat{k}_{0} + \hat{k}_{1}\Delta t$ is typically used to track its frequency drift over time, where $\hat{k}_{0}$ is an estimate of the maser frequency at time $t$ and $\hat{k}_{1} = \frac{\textrm{d}f}{\textrm{d}t}(t)$ is an estimate of its linear frequency drift rate.

For the OLO, a linear model is sufficient to form a time scale with competitive performance.  However, the presence of random walk frequency noise and the exponential term in its frequency drift tend to add a slight curvature to the OLO frequency evolution.  We found that it was more optimal to model the OLO frequency using a quadratic function of the form $f(t+\Delta t) = \hat{k}_{0} + \hat{k}_{1}\Delta t + \frac{1}{2}\hat{k}_{2}\Delta t^{2}$, where the additional term represents the estimate of the time derivative of the linear drift rate.  As shown below, this function is typically written in matrix form, and the coefficients $\hat{k}_0, \hat{k}_1,$ and $\hat{k}_2$ form what is known as the state estimate vector.

For the analysis presented in Fig. 3 the time separation between updates of the Kalman filter state vector was $\Delta t = 1$ s.  For the nth epoch, the local oscillator frequency at time $t_{n}$ is predicted using the state estimate vector at $t_{n-1}$ and its expected evolution over the interval $\Delta t$.  

\begin{equation}\tag{S1}
    \begin{pmatrix} 
    \hat{k}_{0}[n|n-1]  \\
    \hat{k}_{1}[n|n-1] \\
    \hat{k}_{2}[n|n-1]
    \end{pmatrix} =     \begin{pmatrix} 
    1 & \Delta t & \frac{1}{2}\Delta t^2\\
    0 & 1 & \Delta t \\
    0 & 0 & 1
    \end{pmatrix} \begin{pmatrix} 
    \hat{k}_{0}[n-1]  \\
    \hat{k}_{1}[n-1] \\
    \hat{k}_{2}[n-1]
    \end{pmatrix}
\end{equation}

\noindent The vector $\textbf{k}[n|n-1]$ is known as the prior estimate of the state at epoch $n$.  The local oscillator frequency is steered by applying a frequency correction of $\hat{k}_0[n|n-1]$ during each epoch.  In the event that no new measurement with the optical clock occurs during this epoch, the new state estimate vector is simply set equal to the prior estimate.  

\begin{equation}\tag{S2}
    \begin{pmatrix} 
    \hat{k}_{0}[n]  \\
    \hat{k}_{1}[n] \\
    \hat{k}_{2}[n]
    \end{pmatrix} = \begin{pmatrix} 
    \hat{k}_{0}[n|n-1]  \\
    \hat{k}_{1}[n|n-1] \\
    \hat{k}_{2}[n|n-1]
    \end{pmatrix}
\end{equation}

This is the behavior of the steering algorithm when the optical clock is offline.  If the clock is operational and a new frequency measurement is available during epoch $n$, we may asses the fidelity of the model prediction by comparing the prior prediction of the local oscillator frequency, $\hat{k}_0[n|n-1]$, with the measurement:

\begin{equation}\tag{S3}
    \Delta f[n] = y[n] - \hat{k}_0[n|n-1]
\end{equation}

\noindent where $y[n]$ is the measurement of the frequency detuning of the free-running local oscillator relative to the atomic transition
and $\Delta f[n]$ is the prediction error (ie the residual detuning after steering the OLO).  If the optical clock is viewed as an ideal frequency reference, $\Delta f[n]$ represents the residual frequency fluctuations of the local oscillator after steering.  As described in the next section, this quantity can be analyzed to estimate the time error of the time scale.  Following the measurement the state vector is updated as follows:

\begin{equation}\tag{S4}
    \begin{pmatrix} 
    \hat{k}_{0}[n]  \\
    \hat{k}_{1}[n] \\
    \hat{k}_{2}[n]
    \end{pmatrix} = \begin{pmatrix} 
    \hat{k}_{0}[n|n-1]  \\
    \hat{k}_{1}[n|n-1] \\
    \hat{k}_{2}[n|n-1]
    \end{pmatrix} + \Delta f[n] \textbf{K}_n
\end{equation}

\noindent The matrix $\textbf{K}_n$ is known as the optimal Kalman filter gain for epoch $n$.  The gain matrix $\textbf{K}_n$ determines the relative weight of the measurement and the prior estimate when computing the new state estimate vector and is optimized using knowledge from past measurements and the known noise properties of the local oscillator.  The procedure for calculating $\textbf{K}_n$ is beyond the scope of the current discussion.  The reader is referred to the canonical text on this subject~\cite{Kalman}.  Nonetheless, we provide some of the parameters for tuning the Kalman filter below as a reference for readers with expertise in Kalman filtering who wish to reproduce the filter used in this work for steering a comparable local oscillator.

When computing the optimal Kalman filter gain, two co-variance matrices must be specified by the user corresponding to process noise and measurement noise.  Process noise represents uncertainty in the future state of the local oscillator.  For our system, this is dominated by the random-walk frequency noise which limits the local oscillator stability at 1-day averaging intervals.  Measurement noise represents uncertainty in the current frequency measurement due to relative noise between the optical clock and the local oscillator.

The process noise in our Kalman filter is characterized by the $Q$ matrix, defined as $[Q_{11},0,0; 0,Q_{22},0; 0,0,Q_{33}]$.  $Q_{11}$ corresponding to the random walk noise of silicon cavity, is set to be $5.1 \times 10^{-36} (s/s)^2$ which is based on the random walk coefficient from the local oscillator noise model in Fig. 4. $Q_{22}$ and $Q_{33}$, corresponding to random run noise and higher-order noise, are negligible. In practice, these two values are set to $2.2 \times 10^{-46} (s/s^2)^2$ and $3.5 \times 10^{-57} (s/s^3)^2$ respectively. The prediction error is not particularly sensitive to the values of $Q_{22}$ and $Q_{33}$.  Only coarse tuning of these parameters is required to guarantee the performance of our filter. The measurement noise is characterized by the $R$ matrix which has a value of $2.5 \times 10^{-33} (s/s)^2$ corresponding to the Allan variance of the thermal noise floor of the local oscillator.  To test the robustness of the Kalman filter, the $R$ matrix parameter was varied by an order of magnitude in either direction and the analysis presented in Fig. 3 was recomputed.  We found that this impacted the peak-to-peak time error over the 34 day campaign by less than 5 percent.

\section{Estimation of the time error}

To minimize the time error, the time scale steering algorithm must accurately predict and correct for changes in the local oscillator frequency over time.  If the exact frequency evolution of the local oscillator were somehow known, one could compute the time error over a given interval by multiplying the mean prediction error during this period by the elapsed time $\Delta \textrm{t}$.  In this work, a $^{87}$Sr lattice clock with low-10$^{-18}$ level systematic uncertainty~\cite{Toby} and mid-10$^{-17}$ level stability at 1\hspace{0.5mm}s~\cite{Eric} serves as a nearly ideal reference for monitoring the frequency of the OLO.  Due to the intermittency of the measurement record, the exact mean fractional error in the Kalman filter prediction during the gaps between measurements is not known.  Instead, the mean prediction error is estimated by averaging the values immediately before and after the gap.  This estimation is then multiplied by the gap duration to compute a time error for the gap in question.  The contributions from each gap are integrated to produce the dark blue trace in Fig. 3.  Immediately following a longer gap, the prediction error may be appreciable due to a discrepancy between the predicted and actual drift of the OLO during the gap.  The prediction error drops rapidly in the subsequent epochs as the state vector is adjusted to compensate according to Eqn. S4.

\begin{figure}[t]

\includegraphics[width=3.37in]{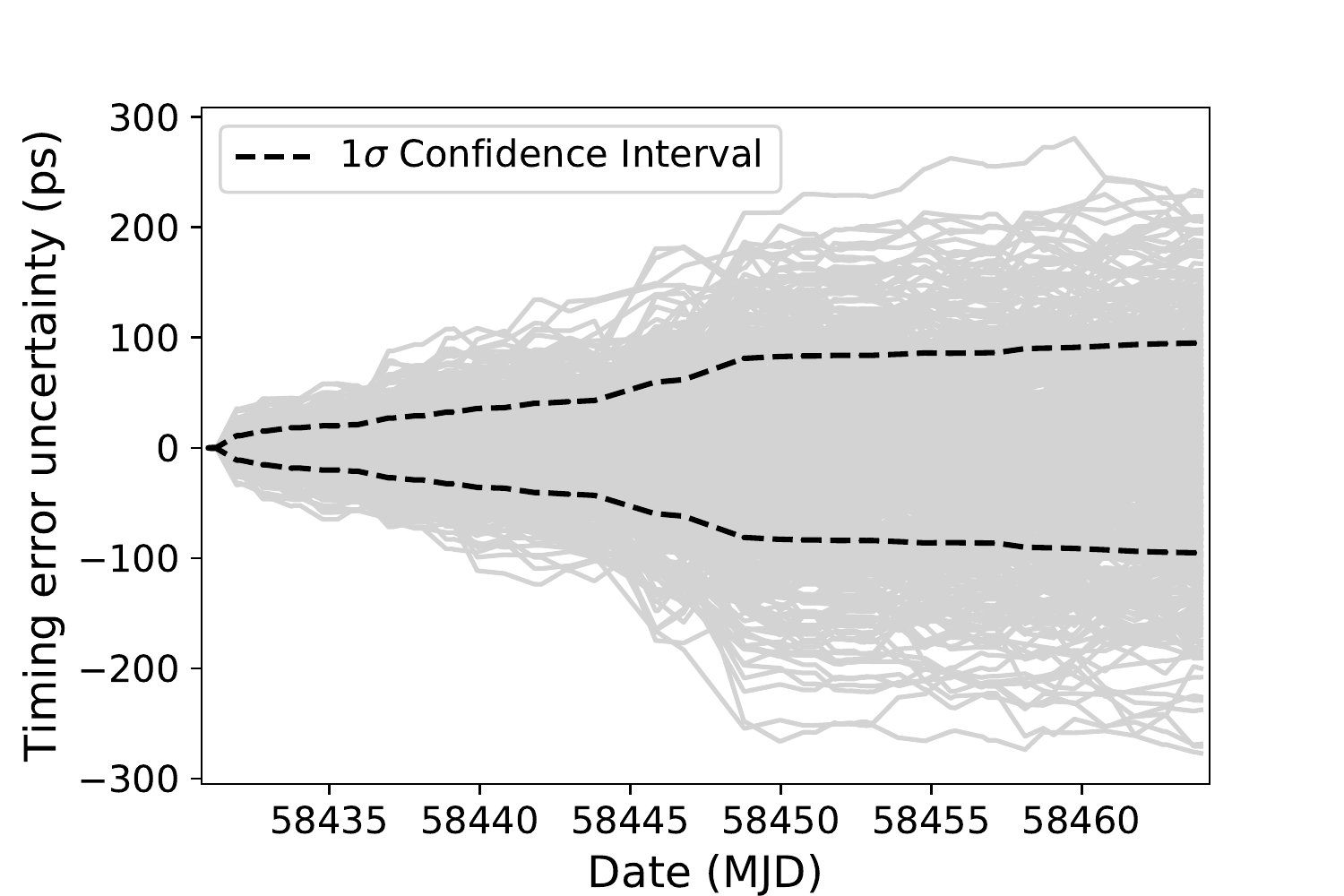}
\caption{Time error uncertainty.  Using the OLO noise model from Fig. 4, we simulate a frequency record for each gap between frequency measurements with the optical clock.  We estimate the uncertainty in the time error calculated in Fig. 3 by computing the difference between the estimated mean frequency and true mean frequency of the OLO during each gap and multiplying by the gap duration.  The outcome of repeated simulations along with a $1\sigma$ confidence interval (dashed line) are depicted.  This corresponds to a $1\sigma$ confidence interval of $\pm94$ ps over the 34 day campaign.}
\label{timeError}
\end{figure}

There is inherent uncertainty in calculating the time error using this method, since the true mean prediction error during a gap likely differs from the estimated value.  To estimate this uncertainty, we assume that during any long gaps between measurements the frequency corrections applied to the OLO by the steering algorithm are only effective in removing the linear plus exponential drift trend shown in Fig. 2(a).  The residuals are assumed to have a frequency stability identical to that of Fig. 2(b) and Fig. 4.  Using the noise model from Fig. 4, we simulate a frequency record for each gap longer than 3 minutes during the 34 day campaign and compute both the mean prediction error for the entire gap and an estimate of the mean prediction error computed by averaging the prediction error for the first and final data points.  By multiplying difference between the true and estimated means by the gap duration, we compute the uncertainty in the time error for the simulation.  By running 1000 simulations, we compute a $1\sigma$ confidence interval for the time error by computing the standard deviation of the simulation result for each gap.  Integrating this over the entire 34 day campaign, we calculate a time error uncertainty of $\pm94$ ps.  The results of each simulation and the resulting $1\sigma$ confidence interval for the integrated time uncertainty are shown in Fig. ~\ref{timeError}.  This $1\sigma$ confidence interval corresponds to the light blue shaded region in Fig. 3.


\section{Expected stability of the optical time scale}

\begin{table}[b]
    \renewcommand\thetable{S1}
    \centering
    \begin{tabular}{c|c|c|c}
          & Si & NIST maser & PTB maser\\\hline
         1 hr/day & $2.4\times 10^{-17}$ & $2.8 \times 10^{-16}$ & $8.7\times 10^{-17}$ \\
         6 hr/day & $1.8\times 10^{-17}$ & $1.2\times 10^{-16}$ & $4.1\times 10^{-17}$ \\
         12 hr/day & $1.2\times 10^{-17}$ & $7.0\times 10^{-17}$ & $2.5\times 10^{-17}$ \\
         20 hr/day & $4.0\times 10^{-18}$ & $3.0\times 10^{-17}$ & $9.9\times 10^{-18}$ \\
    \end{tabular}
    \caption{Expected time scale instability after 34 days ($3\times10^{6}$ s) of averaging based on the Dick effect limit for steering each oscillator for different optical clock uptimes. This assumes that the uptime is grouped into a single run per day. }
    \label{tab:1run}
    \vspace{3mm}
    \renewcommand\thetable{S2}

    \begin{tabular}{c|c|c|c}
          & Si & NIST maser & PTB maser\\\hline
         12 hr runs (1 run/day) & $1.2\times 10^{-17}$ & $7.0 \times 10^{-17}$ & $2.5\times 10^{-17}$ \\
         6 hr runs (2 runs/day) & $6.2\times 10^{-18}$ & $6.6\times 10^{-17}$ & $2.0\times 10^{-17}$ \\
         3 hr runs (4 runs/day) & $3.3\times 10^{-18}$ & $6.1\times 10^{-17}$ & $1.8\times 10^{-17}$ \\
         1 hr runs (12 runs/day) & $1.3\times 10^{-18}$ & $4.1\times 10^{-17}$ & $1.6\times 10^{-17}$ \\
    \end{tabular}
    \caption{Expected time scale instability after 34 days ($3\times10^{6}$ s) of averaging based on the Dick effect limit for steering each oscillator. We assume a total uptime of 12 hours/day distributed over a varying number of evenly spaced clock measurements per day.  The all-optical time scale (Si) shows a significant improvement with increasing measurement frequency.}
    \label{tab:manyrun}

\end{table}

To determine the anticipated stability of the time scale, we simulate a lengthy frequency record for the OLO based on the noise model in Fig. 4 and the drift trend from Fig. 2(a).  This record is then steered using the Kalman filter described in Eqns. S1 through S4.  To simulate the clock uptime, the state vector is updated according to Eqn. S2 during epochs when the clock is offline and updated according to Eqn. S4 when the clock is running.  Unlike the case of a real data set, the true frequency of the local oscillator is known at all times.  Therefore, the prediction error in Eqn. S3 may be computed for every epoch, not just those when the clock is operating.  This prediction error is simply the residual frequency noise of the simulated local oscillator after being steered to the clock.  We then determine the expected fractional frequency stability of the time scale by taking an Allan deviation of these residuals.

\begin{figure}[t]

\includegraphics[width=3.30in]{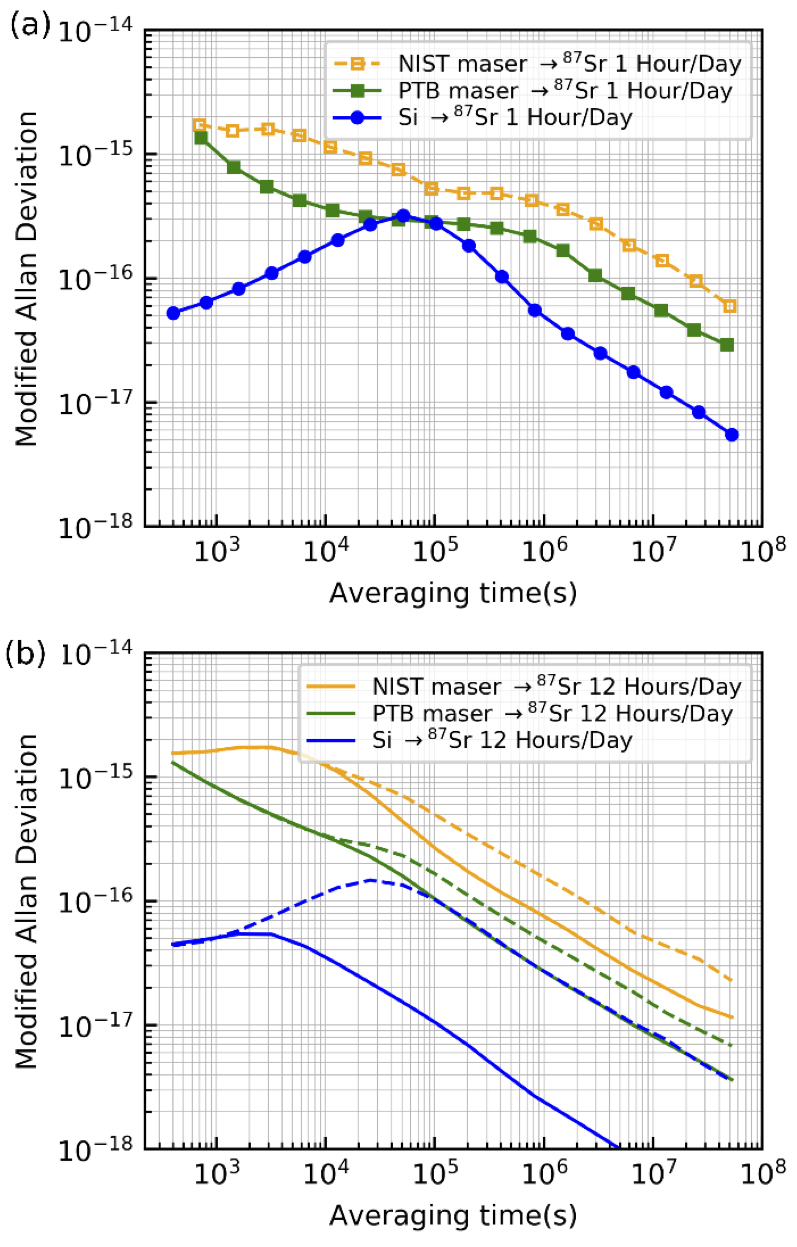}
\caption{Anticipated fractional frequency stability of three different local oscillators steered to an optical clock with several uptime configurations.  Panel (a) assumes a single 1 hour clock measurement per day.  The dashed lines in panel (b) show the stability for a single 12 hour run per day, while the solid line assumes twelve 1 hour runs that are evenly spaced throughout the day.}
\label{SteeringStability}
\end{figure}

Fig. 5 in the main text compares the anticipated stability of our all-optical timescale with that of a single Hydrogen maser from the UTC(NIST) time scale steered daily to an optical clock.  Here, we extend this analysis by also considering time scale local oscillators from other metrology institutes~\cite{NICT, PTBTimescale}. To explore the applicability of our OLO to these time scales, we repeat the analysis from Fig. 5 using an oscillator noise model that is representative of the Hydrogen masers in the UTC(PTB) time scale~\cite{PTBTimescale}.

For the simulations in Fig. 5, the clock is operational for the same time window each day.  In this case it is straightforward to compute the expected stability limit at long averaging times due to aliased noise from the local oscillator.  This noise contribution, colloquially known as the Dick effect limit, is a well studied consequence of steering a local oscillator to an atomic reference in the presence of dead time~\cite{dick1987local,yao2018}.  

In Table~\ref{tab:1run}, we consider the expected time scale stability when steering the three local oscillators assuming a single measurement per day ranging in length from 1 to 20 hours.  The Dick effect limit for the OLO is more than a factor of two better for all run configurations.  As the reliability of optical clocks continues to improve, we anticipate that both the total uptime and the frequency of clock measurements will increase.  Table~\ref{tab:manyrun} shows the expected performance for a 50\% uptime when the clock is run multiple times per day at evenly spaced intervals.  Increasing the measurement frequency results in a modest improvement for the two masers and a dramatic improvement for the OLO.  This is a consequence of the local oscillator being limited by random walk frequency noise ($1/f^2$ slope in the frequency domain) at the Fourier frequencies that contribute to the Dick effect.  The Dick limit, which depends on the local oscillator frequency noise at harmonics of the measurement frequency, falls off rapidly when more measurements are performed per day.




The results of time scale simulations for each oscillator are plotted in Fig ~\ref{SteeringStability}.  Fig ~\ref{SteeringStability}(a) shows the anticipated time scale stability for a single 1 hour run per day.  Fig ~\ref{SteeringStability}(b) depicts the case of 50\% uptime.  Both the stability for a single 12 hour run per day and twelve evenly spaced 1 hour runs are shown for each oscillator.  The stability of each simulation past $10^6$ s agree well with the anticipated Dick effect limit from Table~\ref{tab:1run} and Table~\ref{tab:manyrun}.

\section{Time error of simulated microwave time scales}

In Fig. 3, the performance of the all-optical time scale is compared against two simulated time scales consisting of a hydrogen maser steered to either a $^{87}$Sr lattice clock or a $^{133}$Cs fountain clock.  The noise model for the hydrogen maser is based on the best performing oscillator in the UTC(NIST) time scale~\cite{yao2018}.  Both the instability and systematic uncertainty of a $^{133}$Cs fountain are large enough to have a significant impact on the performance of the time scale over 34 days.  The instability assumed for the fountain is identical to that in~\cite{yao2018}, while the systematic uncertainty is assumed to be $1.71\times 10^{-16}$~\cite{Weyers_2018}.

We repeatedly simulate a 34 day frequency record for the hydrogen maser and steer it to a simulated record of either a $^{87}$Sr lattice clock or a $^{133}$Cs fountain clock that accounts for their respective instability.  The same Kalman filter described in~\cite{yao2018} is used for the steering algorithm.  To account for the systematic uncertainty of the fountain clock a constant fractional frequency offset chosen from a normal distribution with a standard deviation equal to the assumed systematic uncertainty of the fountain is added to the steered frequency record.  The resulting fractional frequency record represents the exact mean fractional frequency offset from an ideal reference for every epoch and may be integrated to compute the time error for each simulation.  We repeat the simulation 800 times and calculate the RMS value of the accumulated time error for each epoch to represent the typical performance of each time scale.  The result is plotted in Fig. 3 in the main text.

\section{Frequency Jump analysis}
As depicted in Fig. 1, the OLO stability at short averaging intervals may also be inferred by measuring against two reference ultrastable lasers and performing a three-cornered hat analysis. The OLO and the reference systems based on a 4 K Si cavity and 40 cm ULE cavity are limited by their respective cavity thermal noise floor of $\sigma = 4.6 \times 10^{-17}$, $8.2 \times 10^{-17}$, and  $1.1 \times 10^{-16}$~\cite{Eric, John, Bishof}. Given their comparable short-term stability, one may easily identify any jumps in the OLO frequency larger than the mid $10^{-16}$ level by examining the time series of its frequency difference with the two reference systems.  When a jump is detected, the predictive model of the OLO can be updated using this information.

\begin{figure}

\includegraphics[width=3.37in]{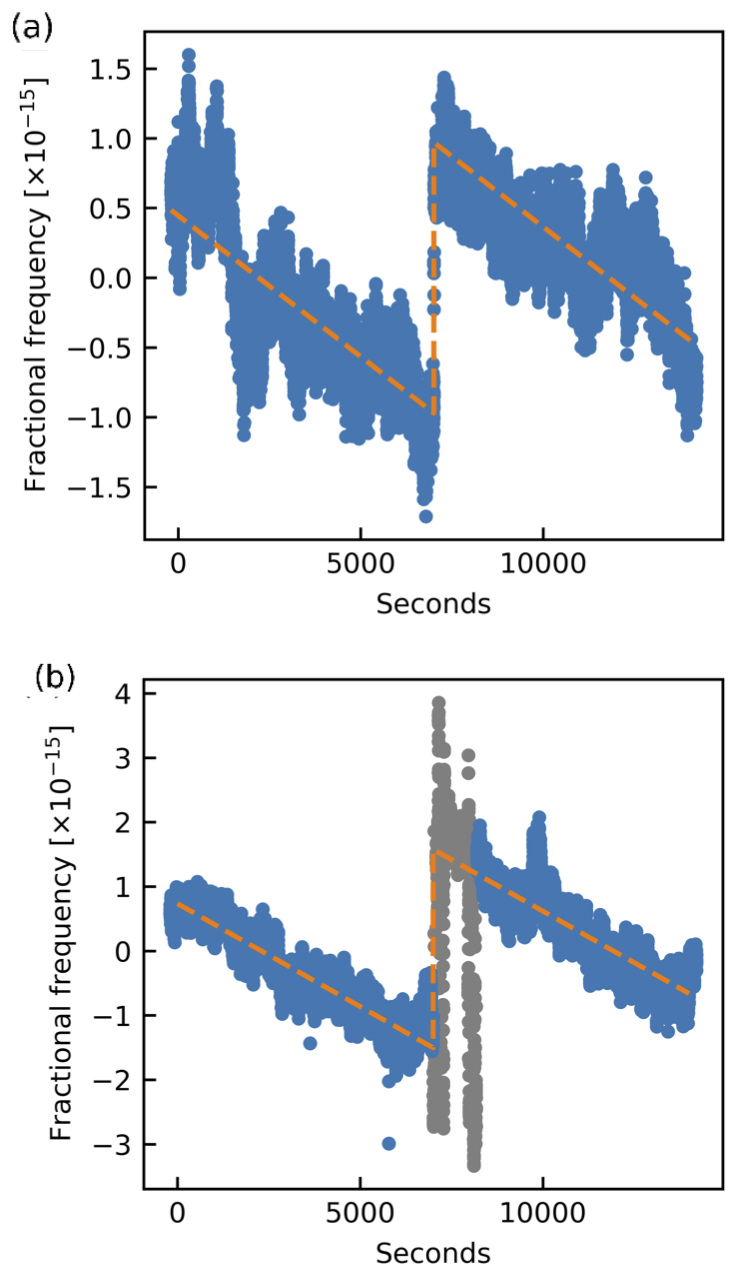}
\caption{Frequency jumps in the difference frequency between the OLO and a reference ultrastable laser based on a 4 K Si cavity. The data are fit to a linear plus Heaviside function to determine the time and magnitude of the jump. The two frequency steps are fit to $1.94 \times 10^{-15}$ and $3.08 \times 10^{-15}$ respectively. Data colored in grey in panel (b) corresponding to the interval when the RAM servo was being debugged is omitted to avoid biasing the fit. }
\label{FrequencyJumpsSi3Si4}
\end{figure}

\begin{figure}

\includegraphics[width=3.37in]{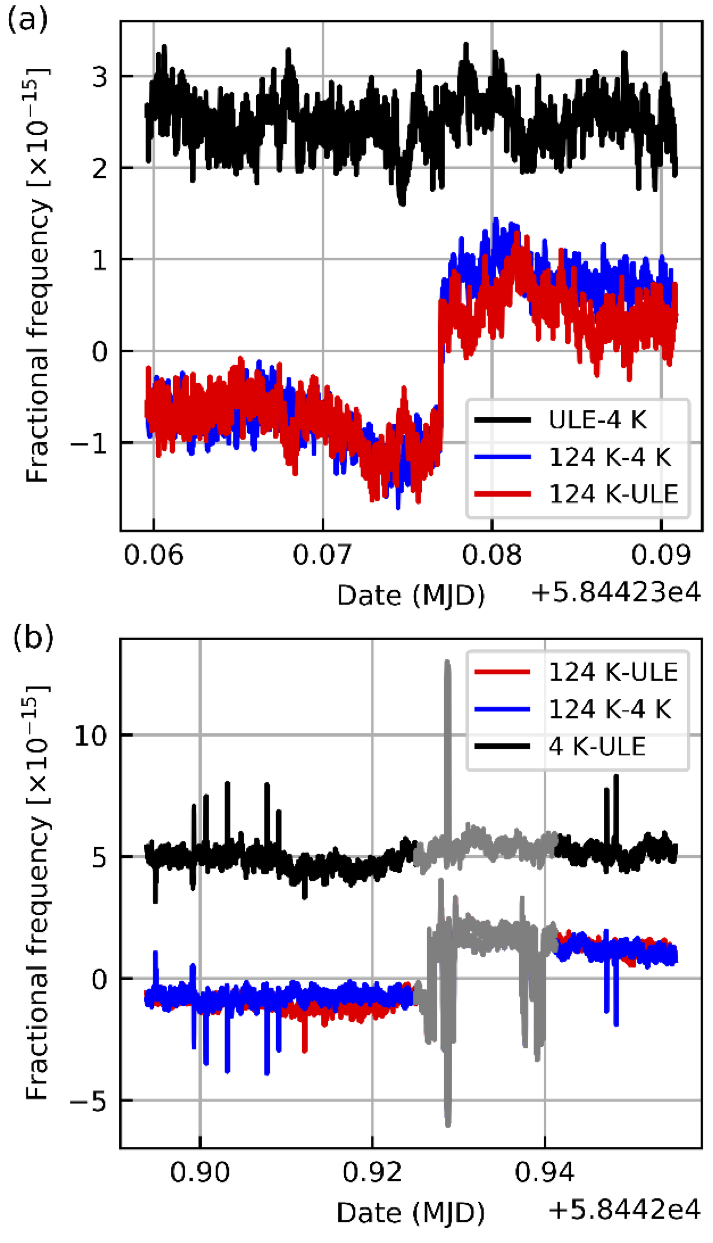}
\caption{Three-cornered hat analysis of the OLO frequency jumps presented in Fig. ~\ref{FrequencyJumpsSi3Si4}.  Examining all three beats, the frequency jumps are clearly attributable to the time scale local oscillator (124 K).}
\label{FrequencyJumpsTCH}
\end{figure}

Between evaluations with the optical clock on MJD 58442 and 58443, two jumps in the OLO frequency were observed using this technique. Fig ~\ref{FrequencyJumpsSi3Si4} displays beat between the OLO and the ultrastable laser based on the 4 K Si cavity, the reference laser with the lowest instability and frequency drift rate.  A 15000 s interval around each frequency jump is fit to a linear plus Heaviside function: $ax + b + d[x > c]$. Here, the c fit coefficient determines the time of the jump in seconds, while the d fit coefficient determines the magnitude of the jump.  Fig~\ref{FrequencyJumpsTCH} shows all three beats during this interval and confirms that both events were frequency jumps on the OLO as the beat between the two reference systems shows no frequency jumps.

The first jump was immediate and has a magnitude in fractional frequency units of $1.94 \times 10^{-15}$. The second jump occurred when the residual amplitude modulation (RAM) servo for the OLO ran out of range.  At this time, we elected to change the sign of the feedback to lock on the opposite side of the error signal.  This allowed us to relock the servo without significantly changing the temperature of the electro-optic modulator as any changes in temperature inside the enclosure housing the ultrastable laser can degrade its stability.  After relocking, an offset in the OLO frequency was observed.  Over a 20 minute interval, the sign of the feedback was changed several times to investigate this effect.  When fitting the frequency step in Fig. \ref{FrequencyJumpsSi3Si4}, only the data outside of this time period is used. The blue data corresponds to the data considered in the fit, while the grey data taken during the locking process was omitted. This second frequency excursion had a magnitude of $3.08 \times 10^{-15}$. A corresponding correction was applied to all data after each frequency step to account for the measured frequency excursion. 

We note that the occurrence of such a jump when the clock is offline can still be corrected for in real time using this approach.  If the jump is identified prior to calculating the Kalman filter prediction for epoch $n$, a constant frequency shift may be added to $\hat{k}_0[n|n-1]$ in Eqn. 1.  The time error introduced by a jump in the frequency of the OLO depends on the time required to identify the jump and apply an appropriate correction to the predictive model to compensate for it.  Due to the excellent short-term stability of the three ultrastable lasers, this can happen almost immediately.  Under the assumption that these two jumps, on average, are identified and corrected for within 1000 s we would anticipate an additional time error of only 5 ps.

\begin{figure}

\includegraphics[width=3.37in]{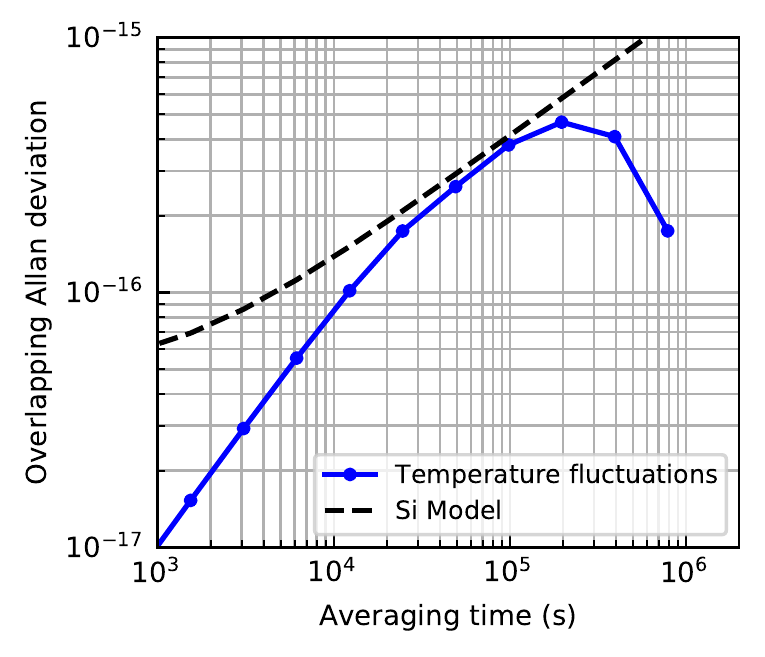}
\caption{Frequency instability arising due to temperature fluctuations assuming an offset of 4.5 mK from the zero-crossing temperature for the Si coefficient of thermal expansion.  The data agrees well with the cavity model from Fig. 4 between $10^4-10^5$s.}
\label{TempFluctuations}
\end{figure}

\section{Optimization of the local oscillator long-term stability}

Parasitic etalons are a known source of instability in optical experiments that have stringent noise requirements at low Fourier frequencies.  Etalon effects can degrade the performance of a number of subsystems used in optical frequency metrology including ultrastable lasers, optical frequency combs, and fiber optic links as well as the sensitivity of advanced interferometric gravitational wave detectors~\cite{Patrick2012,Eric2014,LIGO}.  To illustrate this concept, we consider the case of laser intensity modulation due to a parasitic etalon, though we note that this effect can also lead to phase noise and residual amplitude modulation (RAM).  

\begin{equation}
    \delta P(t) = P_\textrm{C}+P_\textrm{E}+2\sqrt{P_\textrm{C}P_\textrm{E}}\cos\left(\frac{2\pi \delta x(t)}{\lambda}\right)
\end{equation}

\noindent Here $P_\textrm{C}$ is the optical power in the carrier, $P_\textrm{E}$ is a portion of the total power that traverses a slightly different path than the carrier due to scattering off a rough optical surface or an undesired back reflection off of a transmissive optic or photodiode, and $\delta x(t)$ is the time-varying path length difference between the two interfering fields. One sees immediately that this effect can be mitigated by reducing either $P_\textrm{E}$ or $\delta x(t)$.  To limit $P_\textrm{E}$ in our optical setup, we used superpolished mirrors and lenses to reduce optical scatter to $< 10$ ppm per surface.  To limit back reflections from photodiodes, Faraday isolators were installed in front of each detector and their optical windows were removed. Active temperature stabilization of our vacuum chamber was also added to improve the stability of the optical path length between the laser and reference cavity.

When characterizing the OLO prior to beginning this experiment, a strong correlation between slow drift in the optical power incident on the cavity and the laser frequency measured with the Sr clock was observed over a week long measurement.  To address this, a photodetector was installed in transmission and a servo was added to stabilize the transmitted optical power.

The sensitivity of the Si reference cavity to temperature fluctuations is minimized by operating as close as possible to the temperature where the coefficient of thermal expansion (CTE) crosses through zero.  Shortly after the completion of the data campaign, a measurement of the zero crossing temperature similar to that described in~\cite{PTB_drift} revealed that we were operating approximately 4.5 mK above the optimal temperature.  Using the measured temperature fluctuations on the innermost cryogenic shield, the thermal transfer function between this shield and the cavity, and the measured offset from the CTE zero crossing, one may estimate the temperature induced frequency instability of the OLO during the data campaign.  As shown in Fig. ~\ref{TempFluctuations}, we estimate a contribution that is roughly consistent with the measured cavity noise term shown in Fig. 4.  However, no clear temporal correlation between the estimated cavity temperature fluctuations and the frequency residuals from Fig. 2b was observed.  Though this analysis is inconclusive, it suggests that operating closer to the CTE zero crossing will likely be required to significantly improve the OLO stability at averaging times which determine the Dick effect limit of the time scale stability ($10^3-10^5$ s).

\bibliography{main_2}